\documentclass[11pt]{article}
\usepackage{amsfonts,amssymb,a4}

\newcommand{\zeq}{\setcounter{equation}{0}}
\newcommand{\qed}{\hfill\rule{3mm}{3mm}}
\newtheorem{lem}{Lemma}
\newtheorem{teo}{Theorem}

\newtheorem{pro}{Proposition}
\makeatletter \@addtoreset{equation}{section} \makeatother
\begin{document}


\voffset=-1.5truecm\hsize=16.5truecm    \vsize=24.truecm
\baselineskip=14pt plus0.1pt minus0.1pt \parindent=12pt
\lineskip=4pt\lineskiplimit=0.1pt      \parskip=0.1pt plus1pt

\def\ds{\displaystyle}\def\st{\scriptstyle}\def\sst{\scriptscriptstyle}


\let\a=\alpha \let\b=\beta \let\ch=\chi \let\d=\delta \let\e=\varepsilon
\let\f=\varphi \let\g=\gamma \let\h=\eta    \let\k=\kappa \let\l=\lambda
\let\m=\mu \let\n=\nu \let\o=\omega    \let\p=\pi \let\ph=\varphi
\let\r=\rho \let\s=\sigma \let\t=\tau \let\th=\vartheta
\let\y=\upsilon \let\x=\xi \let\z=\zeta
\let\D=\Delta \let\F=\Phi \let\G=\Gamma \let\L=\Lambda \let\Th=\Theta
\let\O=\Omega \let\P=\Pi \let\Ps=\Psi \let\Si=\Sigma \let\X=\Xi
\let\Y=\Upsilon



\def\\{\noindent}
\let\io=\infty

\def\VU{{\mathbb{V}}}
\def\ED{{\mathbb{E}}}
\def\GI{{\mathbb{G}}}
\def\Tt{{\mathbb{T}}}
\def\C{\mathbb{C}}
\def\LL{{\cal L}}
\def\RR{{\cal R}}
\def\SS{{\cal S}}
\def\NN{{\cal M}}
\def\MM{{\cal M}}
\def\HH{{\cal H}}
\def\GG{{\cal G}}
\def\PP{{\cal P}}
\def\AA{{\cal A}}
\def\BB{{\cal B}}
\def\FF{{\cal F}}
\def\TT{{\cal T}}
\def\v{\vskip.1cm}
\def\vv{\vskip.2cm}
\def\gt{{\tilde\g}}
\def\E{{\mathcal E} }
\def\I{{\rm I}}
\def\0{\emptyset}
\def\xx{{\V x}} \def\yy{{\bf y}} \def\kk{{\bf k}} \def\zz{{\bf z}}
\def\ba{\begin{array}}
\def\ea{\end{array}}  \def \eea {\end {eqnarray}}\def \bea {\begin {eqnarray}}

\def\tende#1{\vtop{\ialign{##\crcr\rightarrowfill\crcr
              \noalign{\kern-1pt\nointerlineskip}
              \hskip3.pt${\scriptstyle #1}$\hskip3.pt\crcr}}}
\def\otto{{\kern-1.truept\leftarrow\kern-5.truept\to\kern-1.truept}}
\def\arm{{}}
\font\bigfnt=cmbx10 scaled\magstep1

\newcommand{\card}[1]{\left|#1\right|}
\newcommand{\und}[1]{\underline{#1}}
\def\1{\rlap{\mbox{\small\rm 1}}\kern.15em 1}
\def\ind#1{\1_{\{#1\}}}
\def\bydef{:=}
\def\defby{=:}
\def\buildd#1#2{\mathrel{\mathop{\kern 0pt#1}\limits_{#2}}}
\def\card#1{\left|#1\right|}
\def\proof{\noindent{\bf Proof. }}
\def\qed{ \square}
\def\reff#1{(\ref{#1})}
\def\eee{{\rm e}}

\title{
The Mayer series of the Lennard-Jones gas: improved bounds for the convergence radius
}
\author{
Bernardo N. B. de Lima$^{1}$ and Aldo Procacci$^1$
\\
\small{$^1$ Departamento de Matem\'atica UFMG}
\small{ 30161-970 - Belo Horizonte - MG
Brazil}
}
\maketitle

\def\be{\begin{equation}}
\def\ee{\end{equation}}
\vskip.5cm

\begin{abstract}
We provide a lower bound for the convergence radius of the Mayer series of the Lennard-Jones gas
which strongly improves on the classical bound obtained by Penrose and Ruelle 1963. To
obtain this result we use   an alternative estimate recently proposed by Morais et al.  (J. Stat. Phys. 2014) for a restricted class of stable and tempered pair
potentials (namely those  which can be written as the sum of a
non-negative potential plus an absolutely integrable  and stable
potential) combined with a method  developed  by Locatelli and Schoen (J.  Glob. Optim. 2002)  for establishing a
lower bound for the minimal interatomic distance between particles interacting via a Morse
potential in a cluster of minimum-energy configurations.

 \end{abstract}

\section{Introduction}
In this note we will consider a system of classical particles
interacting via a Lennard-Jones pair potential $V_{\rm LJ}$ (see definition
ahead, formula (\ref{ljtrue})). Such a system is considered as a topical model for a
monoatomic or molecular gas of particles in  statistical mechanics
and the literature about it is  huge. The rigorous approach to
this system has been done mainly in the grand-canonical ensemble. In
this ensemble it is possible to prove that  the pressure of the gas,
which is proportional to  the logarithm of the grand partition
function divided by the volume,  is an analytic function of the
particle fugacity $\l$ as long as $\l\in \mathbb{C}$ is inside a disk of
radius $R$ depending on the inverse temperature $\b$ (and going to $\infty$ as
$\beta\to 0$) uniformly in the volume. This result is physically
interpreted by saying that for such values of $\l$ inside the
convergence region there are no phase transitions and the system is
in the pure gas phase. The best estimate for the   convergence
radius of the pressure series for such systems dates back to the sixties
and has been given independently by Penrose \cite{Pe1,Pe2} and Ruelle
\cite {Ru1,Ru2}. Actually, the Penrose-Ruelle estimate works for a wide
 class of pair potentials, i.e. stable and tempered pair
potential, which of course includes the Lennard-Jones pair potential.
Improvements on the Penrose-Ruelle bound have been recently  given for some specific  cases of stable and tempered pair potentials:
namely,  for  purely hard core gases \cite{FPS} and for  gases interacting via hard-core potentials with an attractive tail \cite{MPS, PU,Pr1,Pr}.
In particular,  an alternative estimate of the same convergence radius has been proposed in \cite{MPS}
for a restricted class of stable and tempered pair
potentials (namely those  which can be written as the sum of a
non-negative potential plus an absolutely integrable  and stable
potential) which  still includes the Lennard-Jones potential. However,
no explicit calculations for the Lennard-Jones potential  are presented in \cite{MPS} due
to the difficulty in evaluating the stability constant of
the absolutely summable part  of the splitting of the pair potential.

In the present paper  we
overcome this difficulty showing  that it is indeed possible to
split the Lennard-Jones potential as described above in such a way
that the stability constant of the absolutely integrable part of the
splitting is exactly equal to the  stability constant of the whole
Lennard-Jones potential.
To obtain this result we use a method  developed  by Locatelli and Schoen in 2002 \cite{LS} for establishing a
lower bound for the minimal interatomic distance between particles interacting via a Morse
potential  (see e.g. \cite{AS,MPS,SABS} for the definition) in a cluster of minimal-energy configurations. The extension of the Locatelli-Schoen technique
to the cut-off Lennard-Jones pair potential
will permit us to efficiently use the new
estimate proposed in \cite{MPS} in such a way that we obtain a lower bound for the convergence radius of the Lennard-Jones gas which
strongly improves the classical one obtained by Ruelle and Penrose in 1963.

The rest of this paper is organized as follows. In Section 2 we will
introduce notations and the model.   We further present our main
results in the form of  a technical lemma (Lemma \ref{lem1}) and, as an immediate corollary of this lemma,  the improvement on the classical bound (Theorem \ref{teo3}).
Finally in Section 3 we present the proof of  Lemma \ref{lem1}
and conclude this section by briefly discussing possible
generalizations.

\section{Notations and Results}
\zeq
\def\xx{{\bf x}}
\def\pp{{\bf p}}
Throughout the paper, if $S$ is a set, then $|S|$ denotes its
cardinality.  If $n\in \mathbb{N}$ is a natural number then we will
denote shortly $\I_n=\{1,2,\dots,n\}$. We also denote by
$\mathbb{Z}^{+}=\mathbb{N}\cup\{0\}=\{0,1,2,\dots\}$ the set of non-negative integers. We  will focus our attention on a system of
classical, identical particles in $\mathbb{R}^3$ (generally enclosed
in a  cubic box $\L\subset \mathbb{R}^3$ with volume $|\L|$). We
denote by $x_i\in \mathbb{R}^3$ the position vector  of the $i^{th}$
particle and by $|x_i|$ its modulus.
We will further assume that these particles interact through a translational and rotational invariant,
stable and tempered pair potential  $V(|x_i-x_j|)$, so that, given $N$
particles in   positions $(x_1,\dots ,x_N)\in \L^N$, their configurational energy $U(x_1,\dots ,x_N)$ is
$$
U(x_1,\dots, x_N)\doteq\sum_{1\le i<j\le N} V(|x_i-x_j|)
$$
We recall that  a pair potential $V(|x_i-x_j|)$ is said to be {\it stable } (see e.g. \cite{Ru}) if
there exists a constant $B\ge 0$ such that
the configurational energy of the $N$
particles in the positions $(x_1,\dots ,x_N)\in \L^N$ satisfies, for all $N\in \mathbb{N}$ and $(x_1,\dots,x_N)\in \mathbb{R}^{3N}$
$$
U(x_1,\dots, x_N)\ge -BN
$$
We also remind the reader that a pair potential $V(|x|)$ is said to be  {\it  tempered} (see again \cite{Ru}) if
there exists a constant $r_0\ge 0$ such that
\be\label{cbetat}
\int_{|x|\ge r_0} |V(|x|)| dx <+\infty
\ee
Moreover, as a consequence of stability and temperedness it is  easy to check that, for all $\beta>0$
\be\label{cbeta}
C(\beta)\doteq \int_{\mathbb{R}^3}
\left|e^{-\b V(|x|)} -1\right|
d x~<~+\infty
\ee
The grand canonical partition function $\Xi_\L(\b,\l)$ of the system is given by
\be\label{n2.2}
\Xi_\L(\b,\l)= 1+ |\L|\l+ \sum_{N\ge 2} {\l^N\over
N!}\int_{\L}\dots\int_{\L}
 e ^{-\b \sum_{1\le i<j\le N}V(|x_i-x_j|) }dx_1\dots dx_N
\ee
with $\b>0$ being the inverse temperature, and $\l>0$ being the fugacity.
The pressure
of the system is  given by
\be\label{pressure}
P(\b,\l)=\lim_{|\L|\to \infty}{1\over\b |\L| }\log \Xi_\L(\b,\l)
\ee

\\The limit (\ref{pressure}) is known to exist whenever the pair potential $V(|x|)$  is stable and tempered   (see e.g. \cite{Ru}, sections 3.3 and 3.4).
Moreover, a very well known and old result (see e.g. \cite{May42,MM, May47, Ru}) states that
 the factor $\log \Xi_\L(\b,\l)$ can be written in terms of a formal series in power of $\l$. Namely,

\be\label{mayers}
{1\over |\L| }\log \Xi_\L(\b,\l)= \sum_{n\ge 1}
C_n(\b,\L){\l^n}
\ee
where $C_1(\b,\L)=1$ and, for $n\ge 2$,
\be\label{urse}
C_n(\b,\L)= {1\over |\L| }{1\over n!}\int_{\L} \dots \int_{\L}
 \sum\limits_{g\in G_{n}} \prod\limits_{\{i,j\}\in E_g}\left[
e^{ -\b V(|x_i -x_j|)} -1\right]dx_1\dots dx_n
\ee
with   $G_n$ being the set of all connected graphs with vertex set
$\I_n$ and $E_g$ denoting the edge-set of a graph $g\in G_n$. The series (\ref{mayers}) is known to converge absolutely for $\l$ sufficiently small
(uniformly in the volume $|\L|$) as long as the pair potential $V(|x|)$ is stable and tempered.
The best rigorous bound on $|C_n(\b,\L)|$ (and consequently on the convergence radius of the series  (\ref{mayers})) to date is the one
obtained by Penrose and (independently) by Ruelle in 1963.
\begin{teo}[Penrose-Ruelle]\label{teo1}
 Let $V(|x|)$ be a stable and tempered  pair  potential. Let $B$ be its stability constant. Then
the  $n$-th order Mayer  coefficient $C_n(\b,\L)$ defined in (\ref{urse})
admits the bound
 \be\label{bmaru}
|C_n(\b,\L)|\le e^{2\b B (n-1)}n^{n-2} {[C(\b)]^{n-1}\over n!}
\ee
where  $C(\b) $ is the function defined in (\ref{cbeta}).

\\Consequently, the Mayer series in the r.h.s. of (\ref{mayers}) converges absolutely, uniformly in $\L$,
 for any complex  $\l$ inside the disk
\be\label{radm}
|\l| <{1\over e^{2\b B+1} C(\b)}
\ee
\end{teo}
\\As said in the introduction, Morais et al. \cite{MPS} proposed  a new   bound  which  can be used in place of the  the Penrose-Ruelle bound  when the pair potentials
 can be written as the sum of non-negative tempered part plus an absolutely summable stable part.

\begin{teo}[Morais-Procacci-Scoppola]\label{teo2}
Let $V(|x|)=\Phi_1(|x|)+\Phi_2(|x|)$ be a pair  potential such that $\Phi_1(|x|)$ is non-negative and tempered  and $\Phi_2(|x|)$ absolutely summable and stable.
Let  $\tilde B$ the stability constant of the potential $\Phi_2(|x|)$.
 Then $n$-th order Mayer coefficient $C_n(\b,\L)$ defined in (\ref{urse})
admits the bound
\be\label{bteo2}
|C_n(\b,\L)|\le e^{{\b \tilde B}n}~n^{n-2} {[\tilde C(\b)]^{n-1}\over n!}
\ee
where
\be\label{tilb}
\tilde C(\b) =\int_{\mathbb{R}^3} \left[ |e ^{-\b \Phi_1(|x|)} -1|+ \b|\Phi_2(|x|)|\right]dx
\ee
Consequently, the Mayer series converges absolutely for all complex activities $\l$  such  that
\be\label{radmru}
|\l|<{1\over e^{{\b \tilde B}+1} \tilde C(\b)}
\ee
\end{teo}
In this note we will focus our attention to the Lennard-Jones pair potential $V_{_{\rm LJ}}(|x|)$. The standard definition in the literature is as follows.
\be\label{ljtrue}
V_{_{\rm LJ}}(|x|)= {1\over |x|^{12}}-{2\over |x|^6}
\ee
Note that the  Lennard-Jones  potential $V_{_{\rm LJ}}(|x|)$ defined in (\ref{ljtrue})
is positive for $|x|<2^{-{1/ 6}}$, negative for $|x|>2^{-{1/ 6}}$ and reaches its minimum at
$|x|=1$ where it takes the value -1.
It is well known that $V_{_{\rm LJ}}(|x|)$  is stable and tempered (see e.g. \cite{FR,Ga,Ru}, i.e. it belongs to the class of pair potentials which satisfy Theorem \ref{teo1}.
Let us denote by  $B_{_{\rm LJ}}$
its stability constant. In other words $B_{_{\rm LJ}}$ is the (minimal) constant such that,
for all $N\in \mathbb{N}$ and $ (x_i,\dots,x_N)\in \mathbb{R}^{3N}$
\be
\sum_{1\le i<j\le N} V_{_{\rm LJ}}(|x_i-x_j|) \ge -B_{_{\rm LJ}}N
\ee
On the other hand,  in \cite{MPS} it has also been shown that $V_{_{\rm LJ}}(|x|)$  belongs to the class of pair potentials satisfying Theorem \ref{teo2} so that one could
also use (\ref{bteo2})-(\ref{radmru}) to get an estimate of the convergence radius. The problem is that
the stability constant of the absolutely summable part of the pair potential  appearing in (\ref{radmru}), being in principle different (and possibly bigger)
than the stability constant $B$  appearing in bound (\ref{radm}),  appears to be quite  difficult to be estimated efficiently.
\vv
\\ In this paper we overcome the difficulties concerning the application of bound (\ref{radmru}) to the  specific case of the Lennard-Jones potential and
we show  that
it is possible to write  $V_{_{\rm LJ}}(|x|)$ as a sum  of a non-negative tempered  part plus an absolutely summable stable part  whose stability constant
is the same constant $B_{_{\rm LJ}}$ of the whole Lennard-Jones potential.  More precisely, we prove the following Lemma

\begin{lem}\label{lem1}
There exists $a\in (0, 2^{-{1\over 6}}) $ such that, defining
\be\label{Valj}
V_a(|x|)= \cases{
{1\over |x|^{12}}-{2\over |x|^6}, &if $|x|>a$\cr\cr
{1\over a^{12}}-{2\over a^6},  & if  $|x|\le a$
}
\ee
the potential $V_a(|x|)$ is stable with stability constant equal to the stability constant of the whole Lennard-Jones potential $V(|x|)$ defined in
(\ref{ljtrue}).
\end{lem}
This Lemma, whose proof  will be given in the next section,
yields straightforwardly new bounds for the convergence radius of the Mayer series of the Lennard-Jones gas which
strongly improve the classical Penrose-Ruelle bound. Indeed,
by writing $V_{_{\rm LJ}}(|x|)= \Phi_1(|x|)+ \Phi_2(|x|)$  with $\Phi_1(|x|)= V_{_{\rm LJ}}(|x|)- V_{a}(|x|)$,
$\Phi_2(|x|)= V_{a}(|x|)$ and choosing the constant $a$ in such way that Lemma \ref{lem1} is satisfied,
we immediately get the following theorem, which is the main result of this note.
\begin{teo}\label{teo3}
Let $V_{_{\rm LJ}}(|x|)$ as in (\ref{ljtrue}). Then there exists  $a\in (0, 2^{-{1\over 6}}) $  such that
the $n$-th order  Mayer coefficient $C_n(\b,\L)$ defined in (\ref{urse})
admits the bound
\be\label{bteo21}
|C_n(\b,\L)|\le e^{{\b  B_{_{\rm LJ}}}n}~n^{n-2} {[\tilde C(\b)]^{n-1}\over n!}
\ee
where
\be\label{tilb2}
\tilde C(\b) =\int_{\mathbb{R}^3} \left[ |e ^{-\b [V_{_{\rm LJ}}(|x|)-V_a(|x|)]} -1|+ \b|V_a(|x|)|\right]dx
\ee
Consequently, the Mayer series  of the Lennard-Jones gas absolutely  converges for all complex activities $\l$  such  that
\be\label{radmru2}
|\l|<{1\over e^{{\b  B_{_{\rm LJ}}}+1} \tilde C(\b)}
\ee
\end{teo}
By (\ref{tilb2}) it is clear that the larger is $a$, the smaller is the factor $\tilde C(\b)$ and consequently
the larger is the lower bound for the convergence radius given by (\ref{radmru2}). We did not try to  optimize $a$, however,
as  shown in the next section, one can take $a= 0.3637$ in order to satisfy Lemma \ref{lem1}.

\\We conclude this section by
giving  an  idea about how  Theorem \ref{teo3} above improves the classical bound of Theorem \ref{teo1}. Putting $\b=1$ for the sake
of simplicity let us denote shortly  $\r_{\rm PR}={1/ (e^{2 B+1} C(\b=1))}$ the Penrose-Ruelle estimate (\ref{radm}) for the convergence radius  and  let
$\r_{\rm new}={1/( e^{ B+1} \tilde C(\b=1))}$ be our new  estimate (\ref{radmru2}) for same the convergence radius.
Observe now that the factor $\tilde C(\b=1)$ appearing in the denominator of  the r.h.s. of (\ref{radmru2}) is surely smaller than 50000. Indeed one can check that

$$
\tilde C(\b=1)\le {4\over 3} \pi (0.3637)^3\left[1+ (0.3637)^{-12}\right] + 4\pi\int_{0.3637}^\infty \left|{1\over r^{12}}-{2\over r^6}\right|r^2dr < 50000
$$
So that we surely can say that
$$
\r_{\rm new}> {e^{-B_{\rm LJ}}\over  50000}
$$
On the other hand  the factor $ C(\b=1)$ appearing in the classical bound (\ref{radm}) is surely larger than 7.89. Indeed it is immediate to check that
$$
C(\b=1)\ge  4\pi\int_{2^{-1/6}}^\infty \left|{1\over r^{12}}-{2\over r^6}\right|r^2dr> 7.89
$$
whence we have that
$$
\r_{\rm PR}<{e^{-2B_{\rm LJ}}\over  7.89}
$$
Thus the ratio $\r_{\rm new}/\r_{\rm PR}$  between the new estimate (\ref{radmru2}) for the convergence radius  and the old estimate (\ref{radm}) for the same convergence radius
is surely larger than
$$
{\r_{\rm new}\over \r_{\rm PR}}> e^{B_{_{\rm LJ}}}{7.89\over 50000} \ge  {e^{B_{_{\rm LJ}}}\over 6338}
$$
The best estimate (from above) for the stability constant $B_{_{\rm LJ}}$ of the Lennard-Jones potential is, as far as we know, the one  recently
obtained  in \cite{SABS}. Namely, $B_{_{\rm LJ}}\le 41.66$. Using this estimate we
have that our new  bound for the convergence radius based on (\ref{radmru}) is,  at inverse temperature $\b=1$,  at least $e^{32.9}$ times large than bound  (\ref{radm}).

\section{Proof of Lemma \ref{lem1}}
\\Let  $a\in (0, 2^{-{1/6}})$ and let $V_a(|x|)$ as in (\ref{Valj}). Given a configuration $(x_1,\dots,x_N)$ of particles
we denote by
$$
U_a(x_1,\dots,x_N)= \sum_{1\le i<j\le N}V_a(|x_i-x_j|)
$$
the  energy of  such configuration. We also  denote (shortly)
$$
W_a(i,\xx_N)= \sum_{j\in \I_N\atop j\neq i} V_a(|x_i-x_j|)
$$
and note that
$$
U_a(x_1,\dots,x_N)= {1\over 2}\sum_{i\in \I_N} W_a(i,\xx_N)
$$
Let now $\xx^*_N\doteq(x^*_1,\dots,x^*_N)$ be  a minimum-energy configuration  for the pair potential  $V_a(|x|)$, i.e. $(x^*_1,\dots,x^*_N)$ is such that
$$
U_a(x^*_1,\dots,x^*_N)\le  U_a(x_1,\dots,x_N)~~~~~~~~~\forall \;\;(x_1,\dots,x_N)\in \mathbb{R}^{3N}
$$
We define
$$
r^a_{min}(\xx^*_N)= \min_{\{i,j\}\subset \I_N}|x^*_i-x^*_j|
$$
We denote by $\mathbb{U}^a_N$ the set of all  minimum-energy configurations of $N$ particles interacting via the pair potential $V_a(|x|)$.

\begin{pro}\label{pro1}
If  $\xx^*_N=(x^*_1,\dots,x^*_N)\in \mathbb{U}^a_N$  then
$$
W_a(i,\xx^*_N)<0,\ \forall i\in I_N
$$
\end{pro}
\\{\bf Proof}. Suppose by contradiction that  $(x^*_1,\dots,x^*_N)\in \mathbb{R}^{3N}$ is a minimum-energy configuration
but $W_a(i,N)\ge 0$. Then we can move the particle $i$ from $x_i^*$ to the position $x_i$
sufficiently far from (the convex hull of) the other particles in
such way that $W_a(i,N)<0$  (recall  that $V_a(|x|)<0$ as soon as
$|x|> 2^{-1/6}$). This new configuration $(x^*_1,\dots,x_i,\dots
x^*_N)$ has thus energy less that the energy of the minimum-energy
configuration $(x^*_1,\dots,x^*_N)$ which is a contradiction. $\Box$
\vv
\vv

\\Let now, for $ (x^*_1,\dots,x^*_N)\in \mathbb{U}^a_N$
$$
W^+_a(i,\xx^*_N)= \sum_{j\in \I_N:\; j\neq i\atop |x^*_i-x^*_j|<2^{-1/6} } V_a(|x^*_i-x^*_j|)
$$
$$
W^-_a(i,\xx^*_N)= \sum_{j\in \I_N:\; j\neq i\atop |x^*_i-x^*_j|\ge2^{-1/6} } V_a(|x^*_i-x^*_j|)
$$
So that
$$
W_a(i,\xx^*_N)=W^+_a(i,\xx^*_N)+W^-_a(i,\xx^*_N)
$$
with
$
W^+_a(i,\xx^*_N)>0$ and $W^-_a(i,\xx^*_N)\le 0.$
Let now define
$$
W^+_a(\xx^*_N)=\max_{i\in I_N} W^+_a(i,\xx^*_N)
$$
Clearly,
for  $\xx_N^*\in \mathbb{U}^a_N$ we have that

\be\label{pro2}
W^+_a(\xx^*_N)\ge V_a(r_{min}(\xx_N^*))
\ee
Indeed, let $k,l\in \I_N$ such that $|x_k^*-x_l^*|=r_{min}(\xx_N^*)$. Then $W^+_a(\xx^*_N)\ge W^+(k,\xx^*_N)\ge V_a(r_{min}(\xx_N^*))$.
\vv

\v\vv

Without loss of generality we assume that the particle with index  $i=1$ is such that $W^+_a(\xx^*_N)= W^+_a(1,\xx^*_N)$
and we also suppose, again without loss of generality, that this particle is in the origin, i.e. $x^*_1=0$.
\\Let now consider, for $n\in \mathbb{N}$, spheres $S_n=\{x\in \mathbb{R}^3:|x|\le 2n\}$ with center in $x^*_1=0$ and radius $2n$ and observe that
\def\vol{{\rm Vol}}
$$
{\rm Vol}(S_n)= {\rm Vol}(S_1) n^3
$$
Let us now define, for  $\xx_N^*\in \mathbb{U}^a_N$,
$$
d_n(\xx_N^*)= {|\{i\in \I_N: x^*_i\in S_n \}|\over \vol(S_n)}
$$
and let
$$
d(\xx^*_N)=\max_{n\ge 1} d_n(\xx_N^*)
$$

\begin{pro}\label{pro3}
Let $\xx_N^*\in \mathbb{U}^a_N$, then there exist a $c_0\in (0,1)$ such that
$$
d(\xx^*_N)\ge c_0{ W^+_a(\xx^*_N)\over {\rm Vol}(S_1)}
$$
\end{pro}

\\{\bf Proof}. First let us recall that, in view of
Proposition \ref{pro1}, we have that
$$
W_a(1,\xx^*_N)=W^+_a(\xx^*_N)+ W^-_a(1,\xx^*_N)<0
$$
and thus
\be\label{1}
W^-_a(1,\xx^*_N)< - W^+_a(\xx^*_N)
\ee
Let now, for any $n\in \mathbb{N}$  (put $S_0=\0$  by convention)
$$
Sol_n=\{i\in I_N: x^*_i\in S_n\setminus S_{n-1}\}
$$
In other words, $Sol_n$ is the set of indices in $I_N$ carried by the particles of the minimum-energy
configuration $\xx_N^*$ which lay between spheres $S_n$ and $S_{n-1}$. Then we have (recall that, for $|x|\ge a$ we have that
$V_a(|x|)= {1\over |x|^{12}}-{2\over |x|^6}$  and the latter is decreasing  for $|x|\ge 1$)
$$
 W^-_a(1,\xx^*_N)\ge - |Sol_1| +\sum_{n=2}^\infty  \left[{1\over (2n-2)^{12}}-{2\over (2n-2)^6}\right]|Sol_n|
$$
Let us first suppose that there exists $n_0\ge 2$ such that
$$
|Sol_{n_0}|\ge n_0^3  W^+_a(\xx^*_N)
$$
then the proposition follows  since
$$
d(\xx^*_N)\ge d_{n_0}(\xx^*_N)= {|\cup_{n=1}^{n_0}Sol_n|\over \vol(S_{n_0})}\ge  {|Sol_{n_0}|\over n_0^3 \vol(S_1)}\ge
{ W^+_a(\xx^*_N)\over \vol(S_1)} \ge c_0 { W^+_a(\xx^*_N)\over \vol(S_1)},\forall c_0\in(0,1)
$$
Let us thus suppose that
$$
|Sol_{n}|< n^3  W^+_a(\xx^*_N)~~~~~~~~~~~~~~{\rm for ~all}~n\ge 2
$$
Then
$$
 W^-_a(1,\xx^*_N)\ge - |Sol_1| +  W^+_a(\xx^*_N)\sum_{n=2}^\infty  \left[{1\over (2n-2)^{12}}-{2\over (2n-2)^6}\right] n^3 \ge
$$
$$
\ge  - |Sol_1| -  {W^+_a(\xx^*_N)\over 32}\sum_{n=2}^\infty {n^3\over (n-1)^6}
$$
We now bound
$$
\sum_{n=2}^\infty {n^3\over (n-1)^6}
\le 8+ {3^3\over 2^6}+{4^3\over 3^6}+\int_4^\infty {x^3\over (x-1)^6}dx<   9
$$
Hence
$$
 W^-_a(1,\xx^*_N)\ge   - |Sol_1| -  {9\over 32} W^+_a(\xx^*_N)
$$
and recalling (\ref{1}) we get
$$
 W^+_a(\xx^*_N) \le   |Sol_1|  + {9\over 32} W^+_a(\xx^*_N)
$$
i.e.
$$
 |Sol_1|\ge  {23\over 32} W^+_a(\xx^*_N)
$$
Observing now that $|Sol_1|= \vol(1) d_1(\xx_N)$ we get that
$$
 d_1(\xx_N) \ge  {23\over 32} {W^+_a(\xx^*_N)\over  \vol(1)}
$$
and the proposition follows by the fact that $d(\xx_N)\ge d_1(\xx_N)$ and we get
\be\label{c0}
c_0\ge{23\over 32}
\ee
$\Box$
\vv

\v\vv
\vv
\\Let us now divide $ \mathbb{R}^3$ into elementary cubes $\D$ of size $\ell$ such that
\be\label{cl}
\sqrt{3}{\ell}<2^{-1/6}
\ee
We suppose
that these cubes $\D$ are half-open and half closed in
an arbitrary way, such that they are disjoint, i.e.
$\D\cap\D'=\emptyset$ if
$\D\neq\D'$, and their union is $\mathbb{R}^3$.   Let us denote by  $\O_n(\ell)$  the number  of elementary cubes necessary to cover  $S_n$. An upper bound for
$\O_n(\ell)$ is
\be\label{bO}
\O_n(\ell)\le \bar\O_n(\ell)\doteq \left({4\over \ell}\right)^3  n^3
\ee
The bound follows by considering that there is a cube of size $4n$ which contains the sphere $S_n$ and this cube contains at most
 $\lceil(4n/ \ell)\rceil^3$ elementary cubes which cover $S_n$.
Observe that
$$
\bar\O_n(\ell)= \bar\O_1(\ell) n^3
$$

\begin{pro}\label{pro4}

For $\xx_N^*\in \mathbb{U}_N^a$,
there exists at least one elementary cube $\D$  containing not less than
\be\label{beta}
\b= \left\lceil c_0{ W^+_a(\xx^*_N)\over|\bar\O_1(\ell)|}\right\rceil
\ee
particles of the minimum-energy configuration $\xx^*_N$.
\end{pro}
\\{\bf Proof}.   Let $n_0$ be the integer  for which $d_n( \xx^*_N)$ is maximal, i.e. $n_0$ is such that
$$
d_{n_0}( \xx^*_N)= d( \xx^*_N)
$$
The number of particles of the   minimum-energy configuration  $\xx_N^*$ inside the sphere $S_{n_0}$ is
$$
d( \xx^*_N) \vol(S_{n_0}) =  n_0^3 d( \xx^*_N)  \vol(S_1)
$$
Consider now the $\bar\O_n(\ell)= \bar\O_1(\ell) n^3$ elementary cubes which surely cover $S_n$. Then one of them must contain a number of particles of the configuration $\xx_N^*$  at least
$$
\left\lceil{d( \xx^*_N) \vol(S_{n_0})\over \bar\O_{n_0}(\ell)}\right\rceil=
\left\lceil{d( \xx^*_N) \vol(S_1)\over \bar\O_1(\ell)}\right\rceil \ge \left\lceil{c_0 W^+_a(\xx^*_N)\over  \bar\O_1(\ell)}\right\rceil
$$
where in the last inequality we use Proposition 3. $\Box$
\vv

\v\vv

\begin{pro}\label{pro5}
Let $\xx_N^*$ be a minimum-energy configuration. Let $\ell\le 0.4275$ and $a\le {\sqrt{3}\over 2}\ell$ then
$$
W^+_a(\xx^*_N) \le  \max\left\{ {V_a\Big({\sqrt{6}\over 2}\ell \Big)\over {c_0\over 2\O_1(\ell)}\left[V_a\Big({\sqrt{6}\over 2}\ell \Big)+ V_a\Big({\sqrt{3}}\ell \Big)\right]-1}~~;~~
{V_a\Big({\sqrt{3}\over 2}\ell \Big)\over {c_0\over 4\O_1(\ell)}\left[V_a\Big({\sqrt{3}\over 2}\ell \Big)+ 3V_a\Big({{{3}\over 2}}\ell \Big)\right]-1}\right\}
$$
\end{pro}
\\{\bf Proof}. In view of Proposition \ref{pro4}, given a minimum energy configuration $\xx_N^*$, there exists an elementary cube $\D$ containing at least $\b$ particles among the $\xx_N^*$.
 Let us subdivide $\D$ in eight subcubes of size
$\ell/2$. Let us consider the pairs of such subcubes whose intersection is just one vertex (opposite pairs). There are two cases to be considered.
\vv
\\1. There is a pair of opposite subcubes such that each subcube of this pair contains at least one particle.

\\2. In any pair of opposite cubes at least one subcube of the pair contains no particle so that all particles are contained in at most 4 subcubes.
\vv
\\{\it Case 1}. Let us assume that there is a pair of opposite subcubes of the cube $\D$ each one containing (at least) a particle. Take two spheres of radius $\sqrt{6}\ell/2$
centered in these two particles.
By a simple geometric argument  it is easy to see that these two spheres cover the whole cube $\D$.
This cube $\D$ contains $\b$ particles,  thus there is at least a particle, say in position $x^*_h$, in one of the two opposite subcubes,
such that at least $\b_1\ge \b/2$ particles of the minimum-energy configuration are at distance less or equal to $\sqrt{6}\ell/2$ from $x^*_h$. Let us denote by $\Sigma$
 the sphere of radius $\sqrt{6}\ell/2$ centered at $x^*_h$. Then, for any particle  $x^*_i$ of the minimum-energy configuration we have
$$
|x^*_h- x^*_i|\le \cases{ {\sqrt{6}\ell\over 2} & if $x^*_i\in \Sigma\cap\D$\cr\cr
\sqrt{3}\ell & otherwise
}
$$
Remember now that the condition (\ref{cl}) for $\ell$ is such that $V_a(\sqrt{3}{\ell})$ is positive. Moreover since by hypothesis $a\le {(\sqrt{3}/2)}\ell$,
we can bound the positive part $W^+_a(h,\xx^*_N)$ of the interaction of the particle $x^*_h$ with all other particles in the minimum-energy configuration as

$$
W^+_a(h,\xx^*_N)= \sum_{j\in \I_N: j\neq h\atop |x_h^*-x_j^*|<2^{-1/6} } V_a(|x_h^*-x_j^*|)\ge (\b_1-1)V_a({\sqrt{6}\ell/2})+ (\b-\b_1)V_a({\sqrt{3}\ell})\ge
$$
$$
\ge  (\b/2-1)V_a({\sqrt{6}\ell/2})+ (\b/2)V_a({\sqrt{3}\ell})
$$
Hence we obtain
$$
W^+_a(\xx^*_N)\ge  {\b\over 2}\left[V_a({\sqrt{6}\ell/2})+ V_a({\sqrt{3}\ell})\right] -  V_a({\sqrt{6}\ell/2})
$$
Recalling now the definition of $\b$ we get
$$
W^+_a(\xx^*_N)\ge  { c_0{ W^+_a(\xx^*_N)\over2|\bar\O_1(\ell)|}}\left[V_a({\sqrt{6}\ell/2})+ V_a({\sqrt{3}\ell})\right] -  V_a({\sqrt{6}\ell/2})
$$
whence
$$
W^+_a(\xx^*_N)\left[ { {c_0 \over2|\bar\O_1(\ell)|}}\left[V_a({\sqrt{6}\ell/2})+ V_a({\sqrt{3}\ell})\right]-1\right] \le V_a({\sqrt{6}\ell/2})
$$
We now have to choose $\ell$ in such way that the l.h.s. of the inequality above is positive. That is, we must impose that
$$
{ {c_0 \over2|\bar\O_1(\ell)|}}\left[V_a({\sqrt{6}\ell/2})+ V_a({\sqrt{3}\ell})\right]>1
$$
i.e, recalling the definitions (\ref{c0}) and (\ref{bO}) of $c_0$  and of  $\bar\O_1(\ell)$ resp. we get the condition
$$
{ {{23\over 32} \over 2   \left({4\over \ell}\right)^3}}\left[{1\over ({\sqrt{6}\ell/2)^{12}}}- {2\over ({\sqrt{6}\ell/2)^{6}}}+
{1\over ({\sqrt{3}\ell)^{12}}}- {2\over ({\sqrt{3}\ell)^{6}}}\right]>1
$$
i.e.
\be\label{l1}
{23\over 2^{12}} \left[{2^6+1\over 3^6}\ell^{-9}- {2\over 3}\ell^{-3}\right]-1>0
\ee
Let us call $\ell_1$ the value of $\ell$ such that the l.h.s. of inequality above is equal to zero. Then one can check that $\ell_1>0.4275$ and hence
(\ref{l1})
is satisfied by taking e.g.
\be\label{c1}
\ell\le 0.4275
\ee
In conclusion, if $\ell$ satisfies (\ref{c1}) and  Case 1 happens, we get that
$$
W^+_a(\xx^*_N)\le {{2^{6}\over  3^6\ell^{12}}- {2^4\over 3^3\ell^{6}}\over {23\over 2^{12}} \left[{2^6+1\over 3^6\ell^{9}}- {2\over 3\ell^{3}}\right]-1}
$$
Note that the function on the r.h.s. restricted to the interval $(0,\ell_1)$ is positive and goes to infinity as $\ell\to 0$ or $\ell\to\ell_1$ and therefore has a minimum
at some point in the interval $(0,\ell_1)$ . By computation one can check that this minimum occurs at a value  slightly greater than $\ell=0.3672$
where the  function takes the value  (slightly less than) 4712
\vv
\\{\it Case 2}. If Case 2 happens then all the $\b$ particles in  the elementary cube $\D$ lie inside four of the eight subcubes.

In that case there is a subcube containing at least $\b_1\ge {\b/4}$ particles.
Choose any particle in this subcube, say the one at position $x^*_h$, then the remaining $\b_1-1$ particles  inside the subcube are all
at distance less or equal $\sqrt{3}\ell/2$ from $x_h^*$ (recall: the subcube has size $\ell/2$). The $\b-\b_1$ particles which are outside
the subcube and inside the elementary cube $\D$ are, in the present Case 2, all contained in a parallepiped
of size $\ell\times \ell\times \ell/2$ which also contains $x^*_h$ and thus they are at distance at most $3\ell/2$ from $x^*_h$. Therefore we
can bound as before
$$
W^+_a(h,\xx^*_N)= \sum_{j\in \I_N:^ j\neq h\atop |x_h^*-x_j^*|<2^{-1/6} } V_a(|x_h^*-x_j^*|)\ge (\b_1-1)V_a({\sqrt{3}\ell/2})+ (\b-\b_1)V_a({{3}\ell/2})\ge
$$
$$
\ge  (\b/4-1)V_a({\sqrt{3}\ell/2})+ (3\b/4)V_a({{3}\ell}/2)
$$
Hence we obtain
$$
W^+_a(\xx^*_N)\ge  {\b\over 4}\left[V_a({\sqrt{3}\ell/2})+ 3V_a({{3}\ell/2})\right] -  V_a({\sqrt{3}\ell/2})
$$
Proceeding as before we now get the inequality
$$
W^+_a(\xx^*_N)\left[ { {c_0 \over 4|\bar\O_1(\ell)|}}\left[V_a({\sqrt{3}\ell/2})+ 3V_a({{3}\ell/2})\right]-1\right] \le V_a({\sqrt{3}\ell/2})
$$
So in Case 2 we must impose that
$$
{ {c_0 \over4|\bar\O_1(\ell)|}}\left[V_a({\sqrt{3}\ell/2})+ 3V_a({{3}\ell}/2)\right]>1
$$
and we get the condition
$$
{ {{23\over 32} \over   4 \left({4\over \ell}\right)^3}}\left[{1\over ({\sqrt{3}\ell/2)^{12}}}- {2\over ({\sqrt{3}\ell/2)^{6}}}+
{3\over ({{3}\ell/2)^{12}}}- {6\over ({{3}\ell/2)^{6}}}\right]>1
$$
i.e.
$$
{23\over 2^{6}} \left[\left({6^5 +2^5\over 3^{11}\ell^{9}}\right)- \left({3^2+1\over 3^5\ell^{3}}\right)\right]-1>0
$$
Let us call $\ell_2$ the value of $\ell$ such that the l.h.s. of inequality above is equal to zero. Then one can check that
$\ell_2\ge 0.6268$ and hence inequality above is widely satisfied if one continues to take
\be\label{c2}
\ell\le \ell_1<0.4275
\ee
In conclusion, if $\ell$ satisfies (\ref{c2}) and Case 2 happens, we get that
$$
W^+_a(\xx^*_N)\le {{2^{12}\over  3^6\ell^{12}}-
{2^7\over 3^3\ell^{6}}\over {23\over 2^{6}} \left[\left({6^5 +2^5\over 3^{11}\ell^{9}}\right)- \left({3^2+1\over 3^5\ell^{3}}\right)\right]-1}
$$
Note that the the function on the r.h.s. restricted to the interval $(0,\ell_2)$ is positive and goes to infinity as $\ell\to 0$ or $\ell\to\ell_2$ and therefore has a minimum
at some point in the interval $(0,\ell_2)$ . By computation one can check that this minimum occurs for (slightly greater than)
 $\ell=0.5385$ where the  function takes the value  (slightly less than) 3020.

\\Therefore we have obtained that
$$
W^+_a(\xx^*_N)\le F(\ell)
$$
where $\ell$ can be any number in the interval $(0,\ell_1)$ and $F(\ell)$ is the function
$$
 F(\ell)=\max\left\{
{{2^{6}\over  3^6\ell^{12}}- {2^4\over 3^3\ell^{6}}\over {23\over 2^{12}} \left[{2^6+1\over 3^6\ell^{9}}- {2\over 3\ell^{3}}\right]-1}~~~;~~~
{{2^{12}\over  3^6\ell^{12}}- {2^7\over 3^3\ell^{6}}\over {23\over 2^{6}} \left[\left({6^5 +2^5\over 3^{11}\ell^{9}}\right)- \left({3^2+1\over 3^5\ell^{3}}\right)\right]-1}\right\}
$$
Note that by the discussion above $F(\ell)$
restricted to the interval $(0,\ell_1)$ is positive and goes to infinity as $\ell\to 0$ or $\ell\to\ell_1$ and therefore $F(\ell)$ has a minimum
in some place in the interval $(0,\ell_1)$. However,
in order to optimize the bound on the convergence radius, it is convenient to choose an $\ell\in (0,\ell_1)$ as large as possible (in such way as to maximize $a$).
By computation one can check that by taking e.g.
$\ell=0.42$  the  function $F(\ell)$ takes the value  (slightly less than) 15545. $\Box$
\vv

\v\vv

\begin{teo}\label{teo4}
If $a= 0.3637$, then the cutoffed Lennard-Jones potential $V_a(|x|)$ defined in (\ref{Valj}) is such that the minimal distance $r^a_{min}$
at which particles in a minimum energy configuration admits the following lower bound
$$
r_{min}\ge 0.44
$$
\end{teo}
\\{\bf Proof}. Choosing $\ell=0.42$,  by Proposition 5, if $a=0.3637< {\sqrt{3}\over 2}(0.42)$ we have that, in any minimum-energy configuration $\xx_N^*$
$$
W^+_a(\xx^*_N)\le F(\ell=0.42)<15545
$$
Since $V_a(r_{min})\le W^+_a(\xx^*_N)$ we obtain
$$
V_a(r_{min})<15545
$$
which yields at least
$
r_{min}\ge 0.446.~
$ $\Box$
\vv

\\We can now conclude the proof of Lemma \ref{lem1}. The  cut-off Lennard-Jones potential $V_a(|x|)$ with $a=0.3637$ is  such that the minimal   distance $r^a_{min}$
between pairs of particles in a minimum-energy configuration  is   greater than 0.44. It is easy to see that this immediately implies that $V_a(|x|)$ is stable
(see e.g. \cite{Ga}, Sec. 4.2).
Moreover  $r^a_{min} > a$. This implies that the stability constant of $V_a(|x|)$ is equal to the stability constant of the whole Lennard-Jones potential $V_{\rm LJ}(|x|)$
(since  the two potentials coincide for distances greater than $a$).
Indeed, any  $x^*_1, \dots, x^*_N$ which is a minimum-energy configuration for $V_a$, since
$V_{\rm LJ}(|x|)\ge  V_a(|x|)$ for all $|x|\le r^a_{min}$ and $V_{\rm LJ}(|x|)=V_a(|x)$ for all $|x|> r^a_{min}$,   is also a minimum-energy configuration for $V$. Viceversa,
since for the whole Lennard-Jones potential $V(|x|)$ the minimal   distance $r_{min}$ between pairs of  particles in a minimum-energy configuration
is also such that $r_{min} > a$  (see e.g. the recent bounds for $r_{min}\ge 0.67985$ obtained in \cite{SABS}),
any  $x^*_1, \dots, x^*_N$ which is a minimum-energy configuration for $V_{\rm LJ}(|x|)$ is also a minimum-energy configuration for $V_a(|x|)$
(since $V_a(|x|)=V_{\rm LJ}(|x|)$ for $|x|\ge r_{min}>a$).

\subsection{Conclusions}
The reasoning developed in this note for the specific case of the Lennard-Jones pair potential (\ref{ljtrue}) could be easily generalized for a physically relevant
class of non-absolutely integrable
pair potentials, namely the so called Lennard-Jones type pair potentials. {We recall that a pair potential $V(|x|)$ is of Lennard-Jones type if there exist $r_0>0$ and $\e>0$ such that}:
$$
V(|x|)\ge {C_1\over|x|^{3+\e}}~~{\rm for}~~ |x|\le r_0, ~~~~
|V(|x|)|\le {C_2\over|x|^{3+\e}}~~{\rm for}~~|x|> r_0
$$
Indeed, Propositions \ref{pro1},  \ref{pro4} and \ref{pro5} are clearly true also for a potential $V(|x|)$ of Lennard-Jones type. On the other hand
Proposition \ref{pro3} holds for a potential $V(|x|)$ of Lennard-Jones type only if $\e>1$.

\section*{Acknowledgments}
The authors have been partially supported by the Brazilian  agencies
Conselho Nacional de Desenvolvimento Cient\'{\i}fico e Tecnol\'ogico
(CNPq) and  Funda{\c{c}}\~ao de Amparo \`a  Pesquisa do Estado de Minas Gerais (FAPEMIG - Programa de Pesquisador Mineiro)

\end{document}